\catcode`\@=11					



\font\fiverm=cmr5				
\font\fivemi=cmmi5				
\font\fivesy=cmsy5				
\font\fivebf=cmbx5				

\skewchar\fivemi='177
\skewchar\fivesy='60


\font\sixrm=cmr6				
\font\sixi=cmmi6				
\font\sixsy=cmsy6				
\font\sixbf=cmbx6				

\skewchar\sixi='177
\skewchar\sixsy='60


\font\sevenrm=cmr7				
\font\seveni=cmmi7				
\font\sevensy=cmsy7				
\font\sevenit=cmti7				
\font\sevenbf=cmbx7				

\skewchar\seveni='177
\skewchar\sevensy='60


\font\eightrm=cmr8				
\font\eighti=cmmi8				
\font\eightsy=cmsy8				
\font\eightit=cmti8				
\font\eightbf=cmbx8				

\skewchar\eighti='177
\skewchar\eightsy='60


\font\ninei=cmmi9
\font\ninesy=cmsy9

\skewchar\ninei='177
\skewchar\ninesy='60


\font\tenrm=cmr10				
\font\teni=cmmi10				
\font\tensy=cmsy10				
\font\tenex=cmex10				
\font\tenit=cmti10				
\font\tensl=cmsl10				
\font\tenbf=cmbx10				
\font\tentt=cmtt10				
\font\tenss=cmss10				
\font\tensc=cmcsc10				
\font\tenbi=cmmib10				

\skewchar\teni='177
\skewchar\tenbi='177
\skewchar\tensy='60

\def\tenpoint{\ifmmode\err@badsizechange\else
	\textfont0=\tenrm \scriptfont0=\sevenrm \scriptscriptfont0=\fiverm
	\textfont1=\teni  \scriptfont1=\seveni  \scriptscriptfont1=\fivemi
	\textfont2=\tensy \scriptfont2=\sevensy \scriptscriptfont2=\fivesy
	\textfont3=\tenex \scriptfont3=\tenex   \scriptscriptfont3=\tenex
	\textfont4=\tenit \scriptfont4=\sevenit \scriptscriptfont4=\sevenit
	\textfont5=\tensl
	\textfont6=\tenbf \scriptfont6=\sevenbf \scriptscriptfont6=\fivebf
	\textfont7=\tentt
	\textfont8=\tenbi \scriptfont8=\seveni  \scriptscriptfont8=\fivemi
	\def\rm{\tenrm\fam=0 }%
	\def\it{\tenit\fam=4 }%
	\def\sl{\tensl\fam=5 }%
	\def\bf{\tenbf\fam=6 }%
	\def\tt{\tentt\fam=7 }%
	\def\ss{\tenss}%
	\def\sc{\tensc}%
	\def\bmit{\fam=8 }%
	\rm\setparameters\setbaselines\fi}


\font\twelverm=cmr12				
\font\twelvei=cmmi12				
\font\twelvesy=cmsy10	scaled\magstep1		
\font\twelveex=cmex10	scaled\magstep1		
\font\twelveit=cmti12				
\font\twelvesl=cmsl12				
\font\twelvebf=cmbx12				
\font\twelvett=cmtt12				
\font\twelvess=cmss12				
\font\twelvesc=cmcsc10	scaled\magstep1		
\font\twelvebi=cmmib10	scaled\magstep1		

\skewchar\twelvei='177
\skewchar\twelvebi='177
\skewchar\twelvesy='60

\def\twelvepoint{\ifmmode\err@badsizechange\else
	\textfont0=\twelverm \scriptfont0=\eightrm \scriptscriptfont0=\sixrm
	\textfont1=\twelvei  \scriptfont1=\eighti  \scriptscriptfont1=\sixi
	\textfont2=\twelvesy \scriptfont2=\eightsy \scriptscriptfont2=\sixsy
	\textfont3=\twelveex \scriptfont3=\tenex   \scriptscriptfont3=\tenex
	\textfont4=\twelveit \scriptfont4=\eightit \scriptscriptfont4=\sevenit
	\textfont5=\twelvesl
	\textfont6=\twelvebf \scriptfont6=\eightbf \scriptscriptfont6=\sixbf
	\textfont7=\twelvett
	\textfont8=\twelvebi \scriptfont8=\eighti  \scriptscriptfont8=\sixi
	\def\rm{\twelverm\fam=0 }%
	\def\it{\twelveit\fam=4 }%
	\def\sl{\twelvesl\fam=5 }%
	\def\bf{\twelvebf\fam=6 }%
	\def\tt{\twelvett\fam=7 }%
	\def\ss{\twelvess}%
	\def\sc{\twelvesc}%
	\def\bmit{\fam=8 }%
	\rm\setparameters\setbaselines\fi}


\font\fourteenrm=cmr12	scaled\magstep1		
\font\fourteeni=cmmi12	scaled\magstep1		
\font\fourteensy=cmsy10	scaled\magstep2		
\font\fourteenex=cmex10	scaled\magstep2		
\font\fourteenit=cmti12	scaled\magstep1		
\font\fourteensl=cmsl12	scaled\magstep1		
\font\fourteenbf=cmbx12	scaled\magstep1		
\font\fourteentt=cmtt12	scaled\magstep1		
\font\fourteenss=cmss12	scaled\magstep1		
\font\fourteensc=cmcsc10 scaled\magstep2	
\font\fourteenbi=cmmib10 scaled\magstep2	

\skewchar\fourteeni='177
\skewchar\fourteenbi='177
\skewchar\fourteensy='60

\def\fourteenpoint{\ifmmode\err@badsizechange\else
	\textfont0=\fourteenrm \scriptfont0=\tenrm \scriptscriptfont0=\sevenrm
	\textfont1=\fourteeni  \scriptfont1=\teni  \scriptscriptfont1=\seveni
	\textfont2=\fourteensy \scriptfont2=\tensy \scriptscriptfont2=\sevensy
	\textfont3=\fourteenex \scriptfont3=\tenex \scriptscriptfont3=\tenex
	\textfont4=\fourteenit \scriptfont4=\tenit \scriptscriptfont4=\sevenit
	\textfont5=\fourteensl
	\textfont6=\fourteenbf \scriptfont6=\tenbf \scriptscriptfont6=\sevenbf
	\textfont7=\fourteentt
	\textfont8=\fourteenbi \scriptfont8=\tenbi \scriptscriptfont8=\seveni
	\def\rm{\fourteenrm\fam=0 }%
	\def\it{\fourteenit\fam=4 }%
	\def\sl{\fourteensl\fam=5 }%
	\def\bf{\fourteenbf\fam=6 }%
	\def\tt{\fourteentt\fam=7}%
	\def\ss{\fourteenss}%
	\def\sc{\fourteensc}%
	\def\bmit{\fam=8 }%
	\rm\setparameters\setbaselines\fi}


\font\seventeenrm=cmr10 scaled\magstep3		


\newdimen\rp@
\newcount\@basestretchnum
\newskip\@baseskip
\newskip\headskip
\newskip\footskip


\def\setparameters{\rp@=.1em
	\headskip=24\rp@
	\footskip=\headskip
	\delimitershortfall=5\rp@
	\nulldelimiterspace=1.2\rp@
	\scriptspace=0.5\rp@
	\abovedisplayskip=10\rp@ plus3\rp@ minus5\rp@
	\belowdisplayskip=10\rp@ plus3\rp@ minus5\rp@
	\abovedisplayshortskip=5\rp@ plus2\rp@ minus4\rp@
	\belowdisplayshortskip=10\rp@ plus3\rp@ minus5\rp@
	\normallineskip=\rp@
	\lineskip=\normallineskip
	\normallineskiplimit=0pt
	\lineskiplimit=\normallineskiplimit
	\jot=3\rp@
	\setbox0=\hbox{\the\textfont3 B}\p@renwd=\wd0
	\skip\footins=12\rp@ plus3\rp@ minus3\rp@
	\skip\topins=0pt plus0pt minus0pt}


\def\setbaselines{\maxdepth=4\rp@\baselinestretch=\@basestretchnum}


\def\baselinestretch{\afterassignment\@basestretch\@basestretchnum}
\def\@basestretch{%
	\@baseskip=12\rp@ \divide\@baseskip by1000
	\normalbaselineskip=\@basestretchnum\@baseskip
	\baselineskip=\normalbaselineskip
	\bigskipamount=\the\baselineskip
		plus.25\baselineskip minus.25\baselineskip
	\medskipamount=.5\baselineskip
		plus.125\baselineskip minus.125\baselineskip
	\smallskipamount=.25\baselineskip
		plus.0625\baselineskip minus.0625\baselineskip
	\setbox\strutbox=\hbox{\vrule height.708\baselineskip
		depth.292\baselineskip width0pt }}



\def\makeheadline{\vbox to0pt{\baselinestretch=1000
	\vskip-\headskip \vskip1.5pt
	\line{\vbox to\ht\strutbox{}\the\headline}\vss}\nointerlineskip}

\def\makefootline{\baselineskip=\footskip\line{\the\footline}}

\def\big#1{{\hbox{$\left#1\vbox to8.5\rp@ {}\right.\n@space$}}}
\def\Big#1{{\hbox{$\left#1\vbox to11.5\rp@ {}\right.\n@space$}}}
\def\bigg#1{{\hbox{$\left#1\vbox to14.5\rp@ {}\right.\n@space$}}}
\def\Bigg#1{{\hbox{$\left#1\vbox to17.5\rp@ {}\right.\n@space$}}}


\mathchardef\alpha="710B
\mathchardef\beta="710C
\mathchardef\gamma="710D
\mathchardef\delta="710E
\mathchardef\epsilon="710F
\mathchardef\zeta="7110
\mathchardef\eta="7111
\mathchardef\theta="7112
\mathchardef\iota="7113
\mathchardef\kappa="7114
\mathchardef\lambda="7115
\mathchardef\mu="7116
\mathchardef\nu="7117
\mathchardef\xi="7118
\mathchardef\pi="7119
\mathchardef\rho="711A
\mathchardef\sigma="711B
\mathchardef\tau="711C
\mathchardef\upsilon="711D
\mathchardef\phi="711E
\mathchardef\chi="711F
\mathchardef\psi="7120
\mathchardef\omega="7121
\mathchardef\varepsilon="7122
\mathchardef\vartheta="7123
\mathchardef\varpi="7124
\mathchardef\varrho="7125
\mathchardef\varsigma="7126
\mathchardef\varphi="7127
\mathchardef\imath="717B
\mathchardef\jmath="717C
\mathchardef\ell="7160
\mathchardef\wp="717D
\mathchardef\partial="7140
\mathchardef\flat="715B
\mathchardef\natural="715C
\mathchardef\sharp="715D


\def\err@badsizechange{%
	\immediate\write16{--> Size change not allowed in math mode, ignored}}

\baselinestretch=1000
\tenpoint

\catcode`\@=12					
\catcode`\@=11
\expandafter\ifx\csname @iasmacros\endcsname\relax
	\global\let\@iasmacros=\par
\else	\immediate\write16{}
	\immediate\write16{Warning:}
	\immediate\write16{You have tried to input iasmacros more than once.}
	\immediate\write16{}
	\endinput
\fi
\catcode`\@=12


\def\rmb{\seventeenrm}

\def\singlespace{\baselineskip=\normalbaselineskip}
\def\halfspace{\baselineskip=1.5\normalbaselineskip}
\def\doublespace{\baselineskip=2\normalbaselineskip}


\def\AB{\bigskip\parindent=40pt
        \centerline{\bf ABSTRACT}\medskip\halfspace\narrower}
\def\AE{\bigskip\nonarrower\doublespace}
\def\nonarrower{\advance\leftskip by-\parindent
	\advance\rightskip by-\parindent}


\def\boxit#1{\vbox{\hrule\hbox{\vrule\kern3pt
	\vbox{\kern3pt#1\kern3pt}\kern3pt\vrule}\hrule}}

\def\hence{\leavevmode\hbox{\bf .\raise5.5pt\hbox{.}.} }

\def\dalemb#1#2{{\vbox{\hrule height.#2pt
	\hbox{\vrule width.#2pt height#1pt \kern#1pt \vrule width.#2pt}
	\hrule height.#2pt}}}
\def\gtorder{\mathrel{\raise.3ex\hbox{$>$}\mkern-14mu
             \lower0.6ex\hbox{$\sim$}}}
\def\ltorder{\mathrel{\raise.3ex\hbox{$<$}\mkern-14mu
             \lower0.6ex\hbox{$\sim$}}}

\newdimen\fullhsize
\newbox\leftcolumn
\def\twoup{\hoffset=-.5in \voffset=-.25in
  \hsize=4.75in \fullhsize=10in \vsize=6.9in
  \def\fullline{\hbox to\fullhsize}
  \let\lr=L
  \output={\if L\lr
        \global\setbox\leftcolumn=\columnbox\global\let\lr=R \advancepageno
      \else \doubleformat \global\let\lr=L\fi
    \ifnum\outputpenalty>-20000 \else\dosupereject\fi}
  \def\doubleformat{\shipout\vbox{
    \fullline{\box\leftcolumn\hfil\columnbox}\advancepageno}}
  \def\columnbox{\leftline{\vbox{\makeheadline\pagebody\makefootline}}}
  \tolerance=1000 }
\twelvepoint
\doublespace
{  \nopagenumbers{
\rightline{~~~October, 2005}
\bigskip\bigskip
\centerline{\rmb Structure of Fluctuation Terms} 
\centerline{\rmb in the Trace Dynamics Ward Identity}           
\medskip                                   
\centerline{\it Stephen L. Adler}
\centerline{\bf Institute for Advanced Study}
\centerline{\bf Princeton, NJ 08540}
\medskip
\bigskip\bigskip
\leftline{\it Send correspondence to:}
\medskip
{\singlespace\leftline{Stephen L. Adler}
\leftline{Institute for Advanced Study}
\leftline{Einstein Drive, Princeton, NJ 08540}
\leftline{Phone 609-734-8051; FAX 609-924-8399; email adler@ias.
edu}  }
\bigskip\bigskip
}}
\vfill\eject
\pageno=2
\AB
We give a detailed analysis of the anti-self-adjoint operator contribution 
to the fluctuation terms in the trace dynamics Ward identity.  
This clarifies the origin of the apparent inconsistency 
between two forms of this identity discussed in Chapter 6 of our recent 
book on emergent quantum theory.  
\AE
\bigskip\bigskip
\vfill\eject
\pageno=3
\centerline{{\bf 1.~~Introduction}}
In our recent book {\it Quantum Theory as an Emergent Phenomenon} [1], we 
developed a classical dynamics of non-commuting matrix (or operator) 
variables, with cyclic permutation inside a trace used as the basic 
calculational tool.  We argued that quantum theory is the statistical 
thermodynamics of this underlying theory, with canonical 
commutation/anticommutation relations, and unitary 
quantum dynamics, both consequences 
of a generalized equipartition theorem. We also argued that fluctuation or  
Brownian motion corrections to this thermodynamics lead to state vector 
reduction and the probabilistic interpretation of quantum theory.  In our 
analysis of fluctuation corrections, we noted that an anti-self-adjoint 
driving term, coming from a self-adjoint contribution to the conserved 
charge $\tilde C$ for global unitary invariance, is needed to give a 
stochastic Schr\"odinger equation that actually reduces the state vector.
However, we also encountered an apparent inconsistency when such an 
anti-self-adjoint driving term was present, in that this term did not 
flip sign appropriately in going from the equation for a fermion operator 
$\psi$ to that for its adjoint $\psi^{\dagger}$.  (See the discussion 
following Eq.~(6.7a) in Chapter 6 of [1].)  

Our aim in this paper is to give a detailed 
analysis of the origin of this apparent inconsistency.  We shall show that 
when details that were glossed over in the treatment of Chapter 6 are taken 
into account, the different forms of the 
Ward identity are always consistent, but in certain cases the 
anti-self-adjoint driving terms tend to cancel.  Specifically, we shall show 
that: (1) A self-adjoint term in $\tilde C$ appears when a fixed 
operator is used in the construction of the fermion kinetic terms, but 
cancels when this operator is elevated to a dynamical variable.  
(2)  In the generic case when a self-adjoint 
term is present in $\tilde C$, the conjugate canonical momentum $p_{\psi}$ 
is no longer equal to $\psi^{\dagger}$. The two equations that are analogs  
of the equations for $\psi$ and $\psi^{\dagger}$ in Eq.~(6.7a) of [1] 
are then 
equations for $\psi$ and $p_{\psi}$, and the fact that the anti-self-adjoint 
driving term has the same sign in both equations is no longer an 
inconsistency.  (3) In special cases where there are degrees of freedom 
with conventional fermion kinetic structure, that couple only indirectly 
through bosonic variables to fermion degrees of freedom that give rise 
to the self-adjoint term in $\tilde C$, the problem noted in 
Chapter 6 of [1] reappears.  However, it is not an inconsistency in the Ward 
identities, but rather an indication that the $\tau $ terms, that 
were neglected 
in the approximations leading to emergent quantum theory, must play a role. 
In other words, in this case, the anti-self-adjoint driving term in the 
stochastic equation cancels to the level of the terms neglected in 
our approximation scheme.

This article is organized as follows.  In Sec. 2 we analyze two models 
for bilinear fermionic Lagrangians, focusing on the structure 
associated with the appearance of a self-adjoint component in $\tilde C$.  
In Sec. 3, we derive the corresponding Ward identities 
analogous to Eq.~(6.7a) of [1].    
In Sec. 4, we discuss the implications of these results for 
the apparent inconsistency discussed in Chapter 6 of [1], leading to the 
conclusions briefly stated above.

\bigskip
\centerline{{\bf 2.~~Analysis of Models for Bilinear Fermionic Lagrangians}}
\bigskip
In this section we analyze two models for bilinear fermionic Lagrangians.  
The first, which generalizes the model developed in Eqs. (2.17) through 
(2.21) of [1], involves a fixed matrix $A_{rs}$ in the fermion kinetic 
term, and develops a self-adjoint contribution to $\tilde C$.  In the 
second, the matrix $A_{rs}$ is elevated to a bosonic dynamical variable, in 
which case its contribution to $\tilde C$ exactly cancels the self-adjoint 
fermionic contribution to $\tilde C$.  

The first model that we consider is based on the bilinear fermionic 
trace Lagrangian 
$${\bf L}= {\rm Tr} \sum_{ra,sa,sb \in F} q_{ra}^{\dagger}A_{rs}
(\dot q_{sa} + q_{sb} B_{ab})+{\rm bosonic}~~~,\eqno(1a)$$
where the notation $\in F$ (which will be suppressed henceforth) 
indicates a sum over fermionic degrees of freedom $q_{ra}$, 
labeled by the composite index $ra$, 
and where the purely bosonic terms are not explicitly shown. Here $A_{rs}$ is a  
fixed bosonic matrix, and $B_{ab}$ is a bosonic operator (a generalized 
gauge potential).  Recalling our 
adjoint convention that for fermionic $\chi_1,\chi_2$, we have 
$(\chi_1\chi_2)^{\dagger}=-\chi_2^{\dagger} \chi_1^{\dagger}$, we see 
that the  Lagrangian of Eq.~(1a) is real up to a total time derivative  
which vanishes in the expression for the trace action, provided that 
$$A_{rs}^{\dagger} = A_{sr}~,~~~ B_{ab}^{\dagger}=-B_{ba}~~~.\eqno(1b)$$
Introducing the canonical momentum defined by 
$$p_{sa}={\delta {\bf L} \over \delta \dot q_{sa} } = \sum_r q_{ra}^{\dagger}
A_{rs}~~~,\eqno(2a)$$
the trace Hamiltonian defined by 
$${\bf H}={\rm Tr} \sum_{sa}p_{sa}\dot q_{sa} - {\bf L}~~~,\eqno(2b)$$ 
has fermionic terms given explicitly by 
$${\bf H}=- {\rm Tr} \sum_{sab} p_{sa}q_{sb}B_{ab}~~~.\eqno(2c)$$
From this we find the equations of motion 
$$\eqalign{
\dot q_{sa}=& -{\delta {\bf H} \over \delta p_{sa} }= -\sum_b q_{sb}B_{ab}
~~~,\cr
\dot p_{sa}=& -{\delta {\bf H} \over \delta q_{sa} }= \sum_b B_{ba}p_{sb}
~~~,\cr
}\eqno(2d)$$
where in the first line we have used the cyclic permutation rule for 
fermionic variables, ${\rm Tr} \chi_1\chi_2= - {\rm Tr} \chi_2 \chi_1$.  

Although the trace Lagrangian in Eq.~(1a) involves the fixed non-commutative 
matrix $A_{rs}$, this does not appear explicitly in the trace Hamiltonian, 
and so the conditions for global unitary invariance of the theory are 
fulfilled.  Consequently, there is a conserved Noether charge $\tilde C$ 
given by 
$$\tilde C=\tilde C_F + \tilde C_B~~~.\eqno(3a)$$
The bosonic part $\tilde C_B$ is given by 
$$\tilde C_B=\sum_{r\in B} [q_r,p_r]~~~,\eqno(3b)$$
and is anti-self-adjoint in the generic case with the bosonic canonical 
variables $q_r,p_r$ either both self-adjoint or both anti-self-adjoint. 
The fermionic part $\tilde C_F$ is given by 
$$\tilde C_F= -\sum_{ra}(q_{ra}p_{ra} + p_{ra} q_{ra})~~~,\eqno(4a)$$
and by using Eq.~(2a) we find that $\tilde C_F$ has a self-adjoint part 
$\tilde C_F^{\rm sa}$ given explicitly by 
$$\tilde C_F^{\rm sa} = {1\over 2} (\tilde C_F + \tilde C_F^{\dagger})
={1\over 2} \sum_{rsa} [A_{rs}, q_{sa}q_{ra}^{\dagger}] ~~~.\eqno(4b)$$
Using the equations of motion of Eq.~(2d), we find that $\tilde C_F$ 
has the time derivative 
$$\dot{\tilde C}_F= -\sum_{rab} [B_{ab}, p_{ra}q_{rb}]
=- \sum_{rsab}[B_{ab},q_{sa}^{\dagger} A_{sr} q_{rb}]~~~,
\eqno(4c)$$ 
from which we see that $\dot{\tilde C}_F$ is anti-self-adjoint, as required 
by the fact that it must cancel against the anti-self-adjoint contribution 
coming from $\dot{\tilde C}_B$.  Thus the self-adjoint part of 
$\tilde C_F$ given in Eq.~(4b) is separately conserved. This can also be  
verified directly by using Eq.~(2d) and its adjoint, together with Eq.~(1b), 
as follows:  
$$\eqalign{
\dot{\tilde C}_F^{\,\rm sa} 
=&{1\over 2} \sum_{rsa} [A_{rs}, \dot{q}_{sa}q_{ra}^{\dagger}
+ q_{sa}\dot{q}_{ra}^{\,\dagger}] \cr
=&-{1\over 2} \sum_{rsab}[A_{rs},q_{sb}B_{ab}q_{ra}^{\dagger}
+q_{sa}B_{ab}^{\dagger}q_{rb}^{\dagger}]\cr
=&-{1\over 2}\sum_{rsab}[A_{rs}, q_{sb}B_{ab}q_{ra}^{\dagger} 
-q_{sa}B_{ba}q_{rb}^{\dagger}] =0~~~.\cr
}\eqno(4d)$$

In writing the Ward identities to be discussed in the next section, several 
auxiliary quantities related to the above discussion will be needed.  First 
of all, we will need a self-adjoint operator Hamiltonian $H$, the trace of 
which gives the trace Hamiltonian ${\bf H} = {\rm Tr} H$. This can be  
constructed from the self-adjoint part of any cyclic permutation of the 
factors in Eq.~(2c), and so is not unique.  We will adopt the simplest 
choice, with fermionic terms given by the expression 
$$H=H^{\dagger} =-{1\over 2}\sum_{sab}(p_{sa}q_{sb}B_{ab} + 
B_{ab}p_{sa}q_{sb})~~~.\eqno(5a)$$
Because this is a function only of the dynamical variables but 
not of the fixed bosonic matrix $A_{rs}$, under a unitary transformation 
of the dynamical variables $p_{sa} \to U^{\dagger} p_{sa} U$, 
$q_{sb} \to U^{\dagger} q_{sb} U$, $B_{ab} \to U^{\dagger} B_{ab} U$, 
the Hamiltonian $H$ of Eq.~(5a) has the attractive feature  
of being unitary covariant, $H \to U^{\dagger}  H U$.  
An alternative expression for the operator Hamiltonian $H$, 
that yields the same trace Hamiltonian ${\bf H}$, is given by
$${1\over 2} \sum_{sab}\left[q_{sb}B_{ab}p_{sa}
+(q_{sb}B_{ab}p_{sa})^{\dagger}\right]
={1\over 2}\sum_{sab}\left[q_{sb}B_{ab}p_{sa}+\sum_{ru}A_{sr}q_{ra}B_{ba}
p_{ub}A^{-1}_{us}\right]~~,\eqno(5b)$$
but since this explicitly involves both $A_{sr}$ and its 
inverse $A^{-1}_{us}$, it is a less natural choice than Eq.~(5a) 
(it is not a unitary covariant, as well as being less tractable),   
and we will not use it in the discussion that follows.

We will also need to evaluate the anticommutator expression 
$$ i_{\rm eff} \tilde C_{\rm eff} \equiv 
{1\over 2}\{ \tilde C,i_{\rm eff} \} \equiv -\hbar (1+{\cal K} + {\cal N})~~~,
\eqno(5c)$$
where $i_{\rm eff}$ and $\hbar$ are the  effective imaginary unit and Planck 
constant given by the ensemble 
expectation $\langle \tilde C \rangle_{\rm AV} = i_{\rm eff} \hbar$ 
(see Eq.~(4.11b) of [1]), and where $-\hbar{\cal K}$ and $-\hbar{\cal N}$ are   
respectively the $c$-number and operator parts of the fluctuating part of 
$i_{\rm eff}\tilde C_{\rm eff}$.  At this 
point we introduce the specialization that the fixed matrix $A_{rs}$ 
commutes with $i_{\rm eff}$, 
$$[i_{\rm eff},A_{rs}]=0~~~,\eqno(6a)$$ 
as a consequence of which, by the cyclic identities, we have  
$${\rm Tr} i_{\rm eff} \tilde C_F^{\rm sa}=
{1\over 2} \sum_{rsa} {\rm Tr} [i_{\rm eff}, A_{rs}] 
q_{sa}q_{ra}^{\dagger}=0~~~.\eqno(6b)$$ 
This implies that it is consistent to ignore the self-adjoint part of 
$\tilde C$ in forming the canonical ensemble.  Since ensemble expectations 
are then functions only of $i_{\rm eff}$, a second consequence of Eq.~(6a) 
is that 
$$\langle \tilde C_F^{\rm sa} \rangle_{\rm AV}=
{1\over 2} \sum_{rsa} [A_{rs},\langle q_{sa} q_{ra}^{\dagger}
\rangle_{\rm AV}]=0
~~~,\eqno(6c)$$
which implies that even in the presence of $\tilde C_F^{\rm sa}$, 
we can still define an effective imaginary unit by the phase of the  
ensemble expectation of $\tilde C$.   

Returning to Eq.~(5c), we  
now specify conditions to make the separation into terms ${\cal K}$ and 
${\cal N}$ unique. In [1] a normal ordering prescription in the emergent 
field theory was invoked, but here we stay within the underlying trace 
dynamics, and impose the natural conditions that ${\cal K}$ and 
${\cal N}$ are respectively the $c$-number part, and the traceless part, 
of Eq.~(5c).  Then as a consequence Eq.~(6b), the self-adjoint part 
$\tilde C_F^{\rm sa}$ makes a vanishing contribution to ${\cal K}$, which 
therefore is a real number, while the operator ${\cal N}$ receives an 
anti-self-adjoint contribution ${\cal N}^{\rm asa}$ given by 
$$-\hbar {\cal N}^{\rm asa}= i_{\rm eff} \tilde C^{\rm sa}_{\rm eff}~~~.
\eqno(6d)$$

Let us turn now to a second model for the bilinear fermionic trace 
Lagrangian, which has a similar structure to that of Eq.~(1a), but with 
the matrix $A_{rs}$ now itself a dynamical variable.  Since $\dot 
A_{rs}$ is no longer zero, to get a trace Lagrangian that is real up to 
time derivative terms, we must redefine the fermion kinetic part of 
Eq.~(1a) according to   
$${\bf L}= {\rm Tr} \sum_{rsab } [q_{ra}^{\dagger}A_{rs}
 (\dot q_{sa} + q_{sb} B_{ab}) 
+ {1\over 2}q_{ra}^{\dagger} \dot A_{rs} q_{sa}] 
+{\rm bosonic}~~~.\eqno(7a)$$
The canonical momentum $p_{ra}$ is unchanged in form, but now there is 
a bosonic canonical momentum $P_{rs}$ conjugate to $A_{rs}$ given by 
$$P_{rs}= {\delta {\bf L} \over \delta \dot A_{rs} }  
= -{1\over 2} \sum_a q_{sa} q_{ra}^{\dagger} + 
 {\delta {\bf L}_{\rm bosonic} \over \delta \dot A_{rs} } ~~~.\eqno(7b)$$ 
Since $(q_{sa}q_{ra}^{\dagger})^{\dagger}= -q_{ra}q_{sa}^{\dagger}$, 
the canonical momentum $P_{rs}$ now has the adjoint behavior  
$P_{rs}^{\dagger} \not= P_{sr}$, and as a consequence, the contribution 
of the canonical pair $A_{rs},P_{rs}$ to $\tilde C$ is no 
longer anti-self-adjoint, but instead has a self-adjoint part 
$$\left(\sum_{rs} [A_{rs},P_{rs}]\right)^{\rm sa} 
={1\over 2} \sum_{rs}[A_{rs}, P_{rs}-P_{sr}^{\dagger}]
= -{1\over 2} \sum_{rsa} [A_{rs}, 
q_{sa} q_{ra}^{\dagger}]~~~, \eqno(7c)$$
which  exactly cancels the self-adjoint fermionic contribution of 
Eq.~(4b).  Thus, when the matrix $A_{rs}$ is elevated to a dynamical 
variable, the Noether charge $\tilde C$ is purely anti-self-adjoint.  

\bigskip
\centerline{\bf 3.~~Fluctuation Terms in the Trace Dynamics Ward Identities}

We proceed now to work out the implications of the Lagrangian of  
Eq.~(1a) for the trace dynamics Ward identities.  To make contact with 
Eqs.~(6.7a) of [1], we shall not need the most general form of these 
identities, but only the statement that the quantities 
${\cal D}q_{ra\,{\rm eff}}$ and ${\cal D}p_{ra\,{\rm eff}}$ vanish when 
sandwiched between general polynomial functions of the ``eff'' projections 
of the dynamical variables, and averaged over the zero source canonical 
ensemble.  The ensemble equilibrium distribution is given by  
$\rho = Z^{-1} \exp(-\lambda {\rm Tr}i_{\rm eff} \tilde C 
-\tau {\bf H})$,   with $\lambda$ and $\tau$ parameters characterizing  
the ensemble, and with  $Z$ (the ``partition function'')  
the ensemble normalizing factor.  For a fermionic $x_u$, the 
expression  ${\cal D}x_{u {\rm eff}}$is given by 
$$\eqalign{
{\cal D} x_{u {\rm eff}} 
=&-\tau \dot x_{u{\rm eff}} {\rm Tr} \tilde C i_{\rm eff} W_{\rm eff} \cr  
+& [i_{\rm eff} W_{\rm eff},  x_{u{\rm eff}}]
+\sum_{s,\ell} \omega_{us} \epsilon_{\ell} \left(W_s^{R\ell} 
{1\over 2}  \{\tilde C,i_{\rm eff} \} 
W_s^{L\ell}\right)_{\rm eff} ~~~.\cr
}\eqno(8)$$
Here  $\omega_{us}$ is a matrix with element -1 when $s$ is the label of 
the variable $x_s$ conjugate to $x_u$, and 0 otherwise, and $W$ is a general 
self-adjoint bosonic polynomial in the dynamical variables.  
The quantities 
in the final term are defined by writing the variation of $W$ when the 
variable $x_s$ is varied (which we denote by 
$\delta_{x_s} W $) in the form 
$$\delta_{x_s} W= 
\sum_{\ell} W_s^{L\ell}\delta x_s W_s^{R\ell}
~~~,\eqno(9a)$$ 
where $\ell$ is a composite index that labels each monomial in the polynomial 
$W$, as well as each occurrence of $x_s$ in the respective  
monomial term.  In this notation we have 
$${\delta {\bf W} \over \delta x_s} =\sum_{\ell} \epsilon_{\ell} W_s^{R\ell} 
W_s^{L\ell}~~~,\eqno(9b)$$
with $\epsilon_{\ell}$ the grading factor 
appropriate to $W_s^{R\ell}$ and to 
$W_s^{L\ell}x_s$ (which must both 
be of the same grade since we have defined $W$ to 
be bosonic).   

We will apply the above expressions when $W$ is taken as the Hamiltonian $H$   
with fermionic terms given by Eq.~(5a). For the fermionic variations   
of $H$ we find 
$$\delta H = -{1\over 2}\sum_{sab} (\delta p_{sa} q_{sb} B_{ab} 
+ p_{sa} \delta q_{sb} B_{ab}
+ B_{ab}  \delta p_{sa} q_{sb} + B_{ab} p_{sa} \delta q_{sb})
~~~,\eqno(10a)$$
from which we can read off the factors $W_s^{L\ell}, W_s^{R\ell}$, and 
$\epsilon_{\ell}$ needed in Eq.~(8).  For example, 
when $x_u$ is the variable $q_{sa}$, the index $s$ in Eq.~(8) labels the 
canonical conjugate variable $p_{sa}$.  Referring to Eq.~(9a), we see  
that the composite index $\ell$ takes the respective values 
1 and $2,b$  for the two factor orderings in Eq.~(10a), with  
$$\eqalign{
W_s^{R1}=&-{1\over 2}\sum_b q_{sb} B_{ab}~,~~W_s^{L1}=1~,~~\epsilon_1=-1
~~~,\cr
W_s^{R2,b}=&-{1\over 2} q_{sb}~~,~~W_s^{L2,b}=B_{ab}~,~~\epsilon_{2,b}=-1
~~~.\cr}\eqno(10b)$$
The corresponding expressions when $x_u$ is the 
variable $p_{sa}$ have a similar structure 
that can be easily read off from the terms in Eq.~(10b) in which 
$\delta q_{sb}$ appears.  Assembling the various pieces of Eq.~(8), and 
using Eqs.~(5c) and (9b), we 
get the following two formulas, 
$$\eqalign{
{\cal D}q_{ra\,{\rm eff}}=&-\tau \dot q_{ra\,{\rm eff}}
{\rm Tr} (\tilde C^{\rm asa}+\tilde C^{\rm sa}) i_{\rm eff} H_{\rm eff}
+i_{\rm eff}[H_{\rm eff},q_{ra\,{\rm eff}}] \cr
-&\hbar(1+{\cal K}) \dot q_{ra\,{\rm eff}}
+{1\over 2} \hbar\sum_b \big( q_{rb}  \{ B_{ab},{\cal N}^{\rm sa} + 
{\cal N}^{\rm asa} \} \big)_{\rm eff}~~~,\cr
{\cal D}p_{ra\,{\rm eff}}=&-\tau \dot p_{ra\,{\rm eff}}
{\rm Tr} (\tilde C^{\rm asa}+\tilde C^{\rm sa}) i_{\rm eff} H_{\rm eff}
+i_{\rm eff}[H_{\rm eff},p_{ra\,{\rm eff}}] \cr
-&\hbar(1+{\cal K}) \dot p_{ra\,{\rm eff}}
-{1\over 2} \hbar\sum_b \big( \{B_{ba}, {\cal N}^{\rm sa} + 
{\cal N}^{\rm asa} \}p_{rb} \big)_{\rm eff}~~~,\cr
}\eqno(11a)$$
where we have explicitly separated $\tilde C$ and ${\cal N}$ into 
self-adjoint (superscript sa) and anti-self-adjoint (superscript asa) parts. 
Taking the adjoint of the first of these equations, and remembering that 
$B_{ab}^{\dagger}=-B_{ba}$, we also get for comparison 
the formula 
$$\eqalign{
({\cal D}q_{ra\,{\rm eff}})^{\dagger}=&-\tau \dot q_{ra\,{\rm eff}}^{\dagger}
{\rm Tr} (\tilde C^{\rm asa}-\tilde C^{\rm sa}) i_{\rm eff} H_{\rm eff}
+i_{\rm eff}[H_{\rm eff},q_{ra\,{\rm eff}}^{\dagger}] \cr
-&\hbar(1+{\cal K}) \dot q_{ra\,{\rm eff}}^{\dagger}
-{1\over 2} \hbar\sum_b \big( \{ B_{ba}, {\cal N}^{\rm sa} - 
{\cal N}^{\rm asa} \} q_{rb}^{\dagger} \big)_{\rm eff}~~~.\cr
}\eqno(11b)$$

\bigskip
\centerline{\bf 4.~~Discussion}

The formulas of Eqs.~(11a) and (11b), which so far involve no approximations,  
are the analogs within the model of Eq.~(1a) of the similar formulas given 
in Eqs.~(6.7a) of ref. [1].  They differ from Eqs.~(6.7a) in a number of 
respects.  
\item{1.}  First of all, the structure of the term involving ${\cal N}$ 
is different from what appears in [1]  
because the simplest choice for the self-adjoint operator 
Hamiltonian $H$, when the 
matrices $A_{rs}$ and $B_{ab}$ are non-trivial operators, has the 
structure of Eq.~(5a), in both terms of which $p_{sa}$ stands to the left  
of $q_{sb}$.  When $B_{ab}=im\delta_{ab}$, corresponding to a mass term,  
this reduces to  $H=-im\sum_{sa}p_{sa}q_{sa}$, which 
when $A_{rs}=\delta_{rs}$ further reduces to $H=-im\sum_{sa} q_{sa}^{\dagger} 
q_{sa}$, which does not have the commutator structure assumed on an 
ad hoc  basis in Eq.~(6.6) of [1].  As a result, in Eq.~(11b) 
the creation operator 
$q_{rb}^{\dagger}$ automatically stands on the right, and the assumption 
made in [1], that ${\cal N}$ is normal ordered, is not necessary.  
\item{2.} In the treatment here, we have based the separation of the 
fluctuation term into ${\cal K}$ and ${\cal N}$ terms on a decomposition 
into $c$-number and traceless parts, rather than an invocation of normal 
ordering.  As a result, we saw that ${\cal K}$ receives no anti-self-
adjoint contribution, and so is a real (rather than a complex) number.   
In terms of the discussion of Chapter 6 of [1], this means that the model 
of Eq.~(1a) does not lead to energy-driven reduction, which requires a 
nonzero imaginary part of ${\cal K}$.  Localization-driven reduction, which 
arises from the anti-self-adjoint part of ${\cal N}$, is still allowed.  
\item{3.}  In the generic case when $A_{rs}$ is not equal to $\delta_{rs}$ 
in any sector, the canonical momentum $p_{sa}$ is not the same as the 
adjoint $q_{sa}^{\dagger}$.   So even when the $\tau$ terms in 
Eqs.~(11a) and (11b) are dropped, there is no contradiction arising from  
the fact that ${\cal N}^{\rm asa}$ appears in the second equation of Eq.~(11a) 
and in Eq.~(11b) with opposite signs.  Thus, in the generic case, the 
inconsistency discussed following Eqs.~(6.7a) of [1] is not present. 
\item{4.} However, there is a specialization of Eqs.~(11a) and (11b) in  
which an analog of the problem noted in ref. [1] persists.  Suppose we 
divide the fermionic degrees of freedom $q_{ra}$ into two classes I and II,  
based on the value of the index $r$, and take $A_{rs}$ to be block diagonal   
within the two classes.  For 
the class I degrees of freedom, we take $A_{rs}$ to be nontrivial, so that 
$p_{ra} \not= q_{ra}^{\dagger}$.  For the class II degrees of freedom, we 
take $A_{rs}=\delta_{rs}$, so that $p_{ra}=q_{ra}^{\dagger}$.  Then if 
we restrict Eqs.~(11a) and (11b) to $r$ values for 
class II degrees of freedom, we see 
that the second equation in Eq.~(11a) has the same structure as Eq.~(11b), 
except that the terms involving  $\tilde C^{\rm sa}$  and 
${\cal N}^{\rm asa}$ both 
have opposite signs in the two equations.  Hence taking the difference 
between the second equation in Eq.~(11a) and Eq.~(11b), we get for 
$r$ in class II, 
$$\eqalign{
{\cal D}q_{ra\,{\rm eff}}^{\dagger} - ({\cal D}q_{ra\,{\rm eff}})^{\dagger}
=&-2\tau \dot q_{ra\,{\rm eff}}^{\dagger}
{\rm Tr} \tilde C^{\rm sa} i_{\rm eff} H_{\rm eff} \cr
-& \hbar\sum_b \big( \{B_{ba}, 
{\cal N}^{\rm asa} \}q_{rb}^{\dagger} \big)_{\rm eff}~~~.\cr
}\eqno(12)$$
This expression must vanish when inserted (sandwiched between polynomials 
in the variables) in canonical ensemble averages. Hence in this case,  
which is a somewhat more general version of the model formulated in 
Eq.~(6.6) of [1], the terms involving ${\cal N}^{\rm asa}$ must effectively 
average to be of the same order of magnitude as the $\tau$ terms, which 
were neglected in the approximation scheme of Chapter 5 of [1].  There 
is no inconsistency in the Ward identities of Eqs.~(11a) and (11b), 
or in Eq.~(12) that is derived from them, but in this case 
one cannot consistently 
drop the $\tau$ terms and reinterpret  these equations as operator equations 
at the level of the emergent quantum theory.  
\bigskip

To conclude, we have reexamined the apparent inconsistency arising from 
Eqs.~(6.7a) of [1], taking into account details that were not sufficiently 
carefully dealt with there. We see that in the generic case one can 
still use operator analogs of Eqs.~(11a) and (11b) as the basis for a state 
vector reduction model.  But we have also seen that there is a tendency 
for the needed anti-self-adjoint driving term to cancel, suggesting 
some caution, and also more speculatively, suggesting a reason why one 
might expect the state vector reduction terms to be small corrections to 
the basic emergent Schr\"odinger equation.  
\bigskip
\centerline{\bf Acknowledgments}
This work was supported in part by the Department of Energy under
Grant \#DE--FG02--90ER40542, and was performed while the author was 
at the Aspen Center for Physics.  
\bigskip
\centerline{\bf References}
\bigskip
\noindent
[1]  Stephen L. Adler, {\it Quantum Theory as an Emergent Phenomenon}, 
Cambridge University Press, 2004. \hfill\break
\bigskip 
\noindent
\vfill
\eject
\bigskip
\bye